\documentclass[prl,twocolumn,superscriptaddress,showpacs]{revtex4}

\usepackage{graphicx}

\begin{document}

\title{Few-Electron Quantum Dot Circuit with Integrated Charge Read-Out}

\author{J. M. Elzerman}
\affiliation{Department of NanoScience and ERATO Mesoscopic
Correlation Project, Delft University of Technology, PO Box 5046,
2600 GA Delft, The Netherlands}
\author{R. Hanson}
\affiliation{Department of NanoScience and ERATO Mesoscopic
Correlation Project, Delft University of Technology, PO Box 5046,
2600 GA Delft, The Netherlands}
\author{J. S. Greidanus}
\affiliation{Department of NanoScience and ERATO Mesoscopic
Correlation Project, Delft University of Technology, PO Box 5046,
2600 GA Delft, The Netherlands}
\author{L. H. Willems van Beveren}
\affiliation{Department of NanoScience and ERATO Mesoscopic
Correlation Project, Delft University of Technology, PO Box 5046,
2600 GA Delft, The Netherlands}
\author{S. De Franceschi}
\affiliation{Department of NanoScience and ERATO Mesoscopic
Correlation Project, Delft University of Technology, PO Box 5046,
2600 GA Delft, The Netherlands}
\author{L. M. K. Vandersypen}
\affiliation{Department of NanoScience and ERATO Mesoscopic
Correlation Project, Delft University of Technology, PO Box 5046,
2600 GA Delft, The Netherlands}
\author{S. Tarucha}
\affiliation{NTT Basic Research Laboratories, Atsugi-shi, Kanagawa
243-0129, Japan} \affiliation{ERATO Mesoscopic Correlation
Project, University of Tokyo, Bunkyo-ku, Tokyo 113-0033, Japan}
\author{L. P. Kouwenhoven}
\affiliation{Department of NanoScience and ERATO Mesoscopic
Correlation Project, Delft University of Technology, PO Box 5046,
2600 GA Delft, The Netherlands}

\date{\today}

\begin{abstract}
We report on the realization of a few-electron double quantum dot
defined in a two-dimensional electron gas by means of surface
gates on top of a GaAs/AlGaAs heterostructure. Two quantum point
contacts (QPCs) are placed in the vicinity of the double quantum
dot and serve as charge detectors. These enable determination of
the number of conduction electrons on each dot. This number can be
reduced to zero while still allowing transport measurements
through the double dot. Microwave radiation is used to pump an
electron from one dot to the other by absorption of a single
photon. The experiments demonstrate that this quantum dot circuit
can serve as a good starting point for a scalable spin-qubit
system.
\end{abstract}
\maketitle

The experimental development of a quantum computer is at present
at the stage of realizing few-qubit circuits. In the solid state,
particular success has been achieved with superconducting devices
in which macroscopic quantum states are used to define two-level
qubit states (see \cite{saclay} and references therein). The
opposite alternative would be the use of two-level systems defined
by microscopic variables, as realized for instance by single
electrons confined in semiconductor quantum dots \cite{curacao}.
For the control of one-electron quantum states by electrical
voltages, the challenge at the moment is to realize an appropriate
quantum dot circuit containing just a single conduction electron.

Few-electron quantum dots have been realized in self-assembled
structures \cite{imamoglu} and also in small vertical pillars
defined by etching \cite{leorep}. The disadvantage of these types
of quantum dots is that they are hard to integrate into circuits
with a controllable coupling between the elements, although
integration of vertical quantum dot structures is currently being
pursued \cite{ono}. An alternative candidate is a system of
lateral quantum dots defined in a two-dimensional electron gas
(2DEG) by surface gates on top of a semiconductor heterostructure
\cite{curacao}. Here, integration of multiple dots is
straightforward by simply increasing the number of gate
electrodes. In addition, the coupling between the dots can be
controlled, since it is set by gate voltages. The challenge is to
reduce the number of electrons to one per quantum dot. This has
long been impossible, since reducing the electron number decreases
at the same time the tunnel coupling, resulting in a current too
small to be measured \cite{ciorga}.

In this report we demonstrate a double quantum dot device
containing a voltage-controllable number of electrons down to a
single electron. We have integrated it with charge detectors that
can read-out the charge state of the double quantum dot with a
sensitivity better than a single electron charge. The importance
of the present circuit is that it can serve as a fully tunable
two-qubit quantum system, following the proposal by Loss and
DiVincenzo \cite{loss}, which describes an optimal combination of
the single-electron charge degree of freedom (for manipulation
with electrical voltages) and the spin degree of freedom (to
obtain a long coherence time).

\begin{figure}[!]
\includegraphics[height=5.8in]{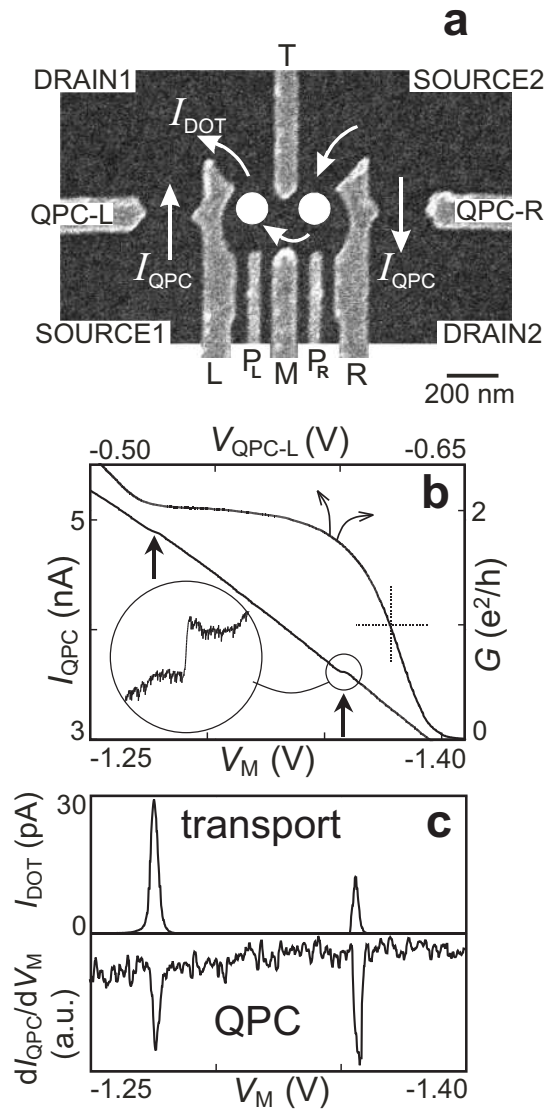}
\caption{\label{Fig 1} (a) Scanning Electron Micrograph of the
metallic surface gates. White circles indicate the two quantum
dots. White arrows show the possible current paths. A bias
voltage, ${V_{DOT}}$, can be applied between source 2 and drain 1,
leading to current through the dots, ${I_{DOT}}$. A bias voltage,
${V_{SD1}}$ (${V_{SD2}}$), between source 1 (source 2) and drain 1
(drain 2), yields a current, ${I_{QPC}}$, through the left (right)
QPC. (b) QPC as a charge detector of the left single dot. Upper
curve with upper and right axis: conductance, ${G}$, of the left
QPC versus the gate voltage, ${V_{QPC-L}}$, showing the last
quantized plateau and the transition to complete pinch-off. The
dashed line indicates the point of highest charge sensitivity.
Lower curve with lower and left axis: current through the left
QPC, ${I_{QPC}}$, versus left-dot gate voltage, ${V_{M}}$.
(${V_{SD1} = 250}$ $\mu$V, ${V_{DOT} = 0}$, ${V_{SD2} = 0}$). The
steps, indicated by the arrows, correspond to a change in the
electron number of the left dot. Encircled inset: the last step
(50 pA high), with the linear background subtracted. (c) Upper
part: Coulomb peaks measured in transport current through the left
dot. Shown is ${I_{DOT}}$ versus ${V_{M}}$ with ${V_{DOT} = 100}$
$\mu$V. Lower part: changes in the number of electrons on the left
dot, measured with the left QPC. Shown is ${dI_{QPC}/dV_{M}}$
versus ${V_{M}}$ (${V_{SD1} = 250}$ $\mu$V, ${V_{DOT} = 0}$).}
\end{figure}

Our device, shown in Fig. 1a, is made from a GaAs/AlGaAs
heterostructure, containing a 2DEG 90 nm below the surface with an
electron density, ${n_{s}=2.9\times 10^{11}}$ cm${^{-2}}$. This
small circuit consists of a double quantum dot and two quantum
point contacts (QPCs). The layout is an extension of previously
reported single quantum dot devices \cite{ciorga}. The double
quantum dot is defined by applying negative voltages to the 6
gates in the middle of the figure. Gate ${T}$ in combination with
the left (right) gate, ${L}$ (${R}$), defines the tunnel barrier
from the left (right) dot to drain 1 (source 2). Gate ${T}$ in
combination with the middle, bottom gate, ${M}$, defines the
tunnel barrier between the two dots. The narrow "plunger" gate,
$P_{L}$ ($P_{R}$), on the left (right) is used to change the
electrostatic potential of the left (right) dot. The left plunger,
$P_{L}$, is connected to a coaxial cable so that we can apply
high-frequency signals. In the present experiments we do not apply
dc voltages to $P_{L}$. In order to control the number of
electrons on the double dot, we use gate ${L}$ for the left dot
and $P_{R}$ for the right dot. All data shown are taken at zero
magnetic field and at a temperature of 10 mK.

We first characterize the individual dots. From standard Coulomb
blockade experiments \cite{curacao} we find that the energy cost
for adding a second electron to a one-electron dot is 3.7 meV. The
excitation energy (i.e. the difference between the first excited
state and the ground state) is 1.8 meV at zero magnetic field. For
a two-electron dot the energy difference between the singlet
ground state and the triplet excited state is 1.0 meV at zero
magnetic field. Increasing the field (perpendicular to the 2DEG)
leads to a transition from a singlet to a triplet ground state at
about 1.7 Tesla.

In addition to current flowing through the quantum dot, we can
measure the charge on the dot using one of the QPCs \cite{field,
sprinzak}. We define only the left dot (by grounding gates ${R}$
and ${P_{R}}$, and use the left QPC as a charge detector. The QPC
is formed by applying negative voltages to \textit{QPC-L} and
${L}$. This creates a narrow constriction in the 2DEG, with a
conductance, ${G}$, that is quantized when sweeping the gate
voltage ${V_{QPC-L}}$. The plateau at ${G = 2e^{2}/h}$ and the
transition to complete pinch-off (i.e. ${G = 0}$) are shown in
Fig. 1b. At the steepest point, where ${G \approx e^{2}/h}$, the
QPC-conductance has a maximum sensitivity to changes in the
electrostatic environment, including changes in the charge of the
nearby quantum dot. As can be seen in Fig. 1b, the QPC-current,
${I_{QPC}}$, decreases when we make the left-dot gate voltage,
${V_{M}}$, more negative. Periodically this changing gate voltage
pushes an electron out of the left dot. The associated sudden
change in charge increases the electrostatic potential in the QPC,
resulting in a step-like structure in ${I_{QPC}}$ (see expansion
in Fig. 1b, where the linear background is subtracted). So, even
without passing current through the dot, ${I_{QPC}}$ provides
information about the charge on the dot. To enhance the charge
sensitivity we apply a small modulation (0.3 mV at 17.7 Hz) to
${V_{M}}$ and use lock-in detection to measure ${dI_{QPC}/dV_{M}}$
\cite{sprinzak}. Figure 1c shows the resulting dips, as well as
the corresponding Coulomb peaks measured in the current through
the dot. The coincidence of the two signals demonstrates that the
QPC indeed functions as a charge detector. From the height of the
step in Fig. 1b (50 pA, typically 1-2 percent of the total
current), compared to the noise (5 pA for a measurement time of
100 ms), we can estimate the sensitivity of the charge detector to
be about ${0.1e}$, with ${e}$ being the single electron charge.
The important advantage of QPC charge detection is that it
provides a signal even when the tunnel barriers of the dot are so
opaque that ${I_{DOT}}$ is too small to measure \cite{field,
sprinzak}. This allows us to study quantum dots even while they
are virtually isolated from the leads.

\begin{figure}[!]
\includegraphics[height=5.2in]{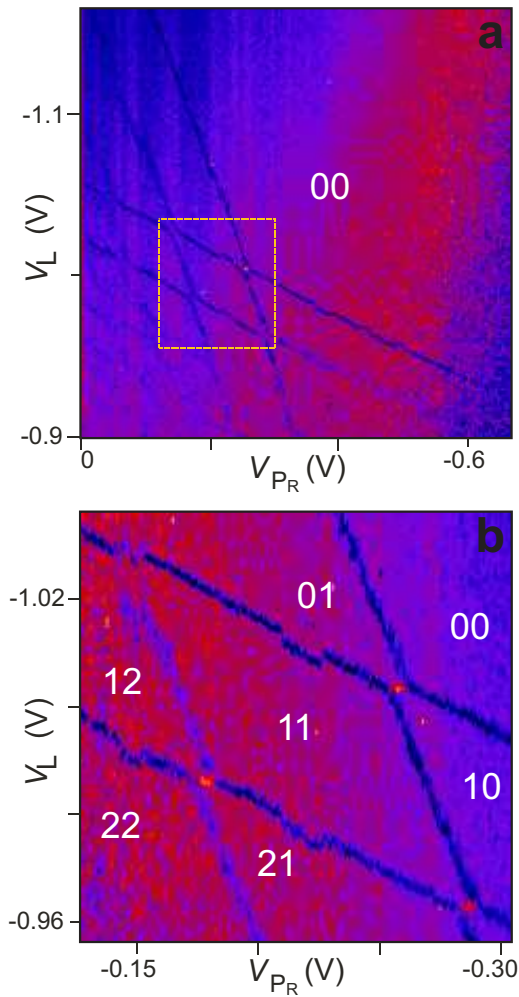}
\caption{\label{Fig 2} (a) Charge stability diagram ("honeycomb")
of the double quantum dot, measured with \textit{QPC-R}. A
modulation (0.3 mV at 17.77 Hz) is applied to gate ${L}$, and
${dI_{QPC}/dV_{L}}$ is measured with a lock-in amplifier and
plotted in color scale versus ${V_{L}}$ and ${V_{PR}}$. The bias
voltages are: ${V_{SD2} = 100}$ $\mu$V and ${V_{DOT} = V_{SD1} =
0}$. The label "00" indicates the region where the double dot is
completely empty. (b) Zoom-in of Fig. 2a, showing the honeycomb
pattern for the first few electrons in the double dot. The white
labels indicate the number of electrons in the left and right
dot.}
\end{figure}

Next, we study the charge configuration of the double dot, using
the QPC on the right as a charge detector. We measure
${dI_{QPC}/dV_{L}}$ versus ${V_{L}}$, and repeat this for many
values of ${V_{PR}}$. The resulting two-dimensional plot is shown
in Fig. 2a. Blue lines signify a negative dip in
${dI_{QPC}/dV_{L}}$, corresponding to a change in the total number
of electrons on the double dot. Together these lines form the
well-known "honeycomb diagram" \cite{pothier, wilfred}. The
almost-horizontal lines correspond to a change in the electron
number in the left dot, whereas almost-vertical lines indicate a
change of one electron in the right dot. In the upper left region
the "horizontal" lines are not present, even though the QPC can
still detect changes in the charge, as demonstrated by the
presence of the "vertical" lines. We conclude that in this region
the \textit{left dot} contains zero electrons. Similarly, a
disappearance of the "vertical" lines occurs in the lower right
region, showing that here the \textit{right dot} is empty. In the
upper right region, the absence of lines shows that here the
\textit{double dot} is completely empty.

We are now able to count the absolute number of electrons. Figure
2b shows a zoom-in of the few-electron region. Starting from the
"00" region, we can label all regions in the honeycomb diagram,
e.g. the label "21" means two electrons in the left dot and one in
the right. Besides the blue lines, also short yellow lines are
visible, signifying a positive peak in ${dI_{QPC}/dV_{L}}$. These
yellow lines correspond to a charge transition between the dots
while the total electron number remains the same. (The positive
sign of ${dI_{QPC}/dV_{L}}$ can be understood if we note that
crossing the yellow lines by making ${V_{L}}$ a little more
positive means moving an electron from the right to the left dot,
which increases ${I_{QPC}}$. Therefore the differential quantity
${dI_{QPC}/dV_{L}}$ displays a positive peak.) The QPC is thus
sufficiently sensitive to detect \textit{inter-dot} transitions.

\begin{figure}[b]
\includegraphics[height=2.4in, clip=true]{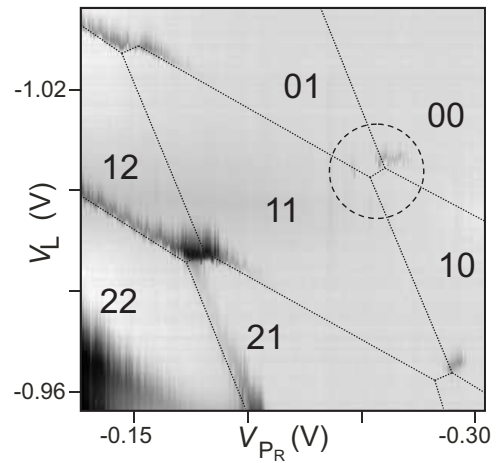}
\caption{\label{Fig 3} Transport through the double dot in the
same region as Fig. 2b. Plotted in logarithmic grayscale is
${I_{DOT}}$ versus ${V_{L}}$ and ${V_{PR}}$, with ${V_{DOT} =
100}$ $\mu$V and ${V_{SD1} = V_{SD2} = 0}$. The dotted lines are
extracted from Fig. 2b. In the light regions current is zero due
to Coulomb blockade. Dark gray indicates current, with the darkest
regions (in the bottom left corner) corresponding to ${\sim 100}$
pA. Inside the dashed circle, the last Coulomb peaks are visible
(${\sim 1}$ pA). (A smoothly varying background current due to a
small leakage from a gate to the 2DEG has been subtracted from all
traces.)}
\end{figure}

In measurements of transport through lateral double quantum dots,
the few-electron regime has never been reached \cite{wilfred}. The
problem is that the gates, used to deplete the dots, also strongly
influence the tunnel barriers. Reducing the electron number would
always lead to the Coulomb peaks becoming unmeasurably small, but
not necessarily due to an empty double dot. The QPC detectors now
permit us to compare charge and transport measurements. Figure 3
shows ${I_{DOT}}$ versus ${V_{L}}$ and ${V_{PR}}$, with the dotted
lines extracted from the measured charge lines in Fig. 2b. In the
bottom left region the gates are not very negative, hence the
tunnel barriers are quite open. Here the resonant current at the
charge transition points is quite high (${\sim100}$ pA, dark
gray), and also lines due to cotunneling are visible [20]. Towards
the top right corner the gate voltages become more negative,
thereby closing off the barriers and reducing the current peaks
(lighter gray). The last Coulomb peaks (in the dashed circle) are
faintly visible (${\sim1}$ pA). They can be increased (up to
${\sim70}$ pA) by readjusting the barrier gate voltages. Apart
from a slight shift, the dotted lines nicely correspond to the
regions where a transport current is visible. We are thus able to
measure transport through a one-electron double quantum dot.

\begin{figure}[!]
\includegraphics[height=2.4in]{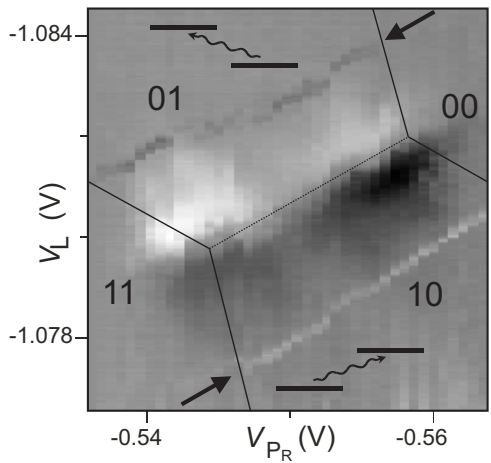}
\caption{\label{Fig 4} Photon-assisted transport through the
double dot, with zero bias voltage, i.e. ${V_{DOT} = V_{SD1} =
V_{SD2} = 0}$. A microwave signal of 50 GHz is applied to
${P_{L}}$. The microwaves pump a current, ${I_{DOT}}$, by
absorption of photons. This photon-assisted current shows up as
two lines, indicated by the two arrows. The white line (bottom)
corresponds to pumping from the left to the right reservoir, the
dark line (top) corresponds to pumping in the reverse direction.
In the middle, around the dotted line, a finite current is induced
by an unwanted voltage drop over the dot, due to asymmetric
coupling of the ac-signal to the two leads \cite{wilfred}.}
\end{figure}

The use of gated quantum dots for quantum state manipulation in
time requires the ability to modify the potential at high
frequencies. We investigate the high-frequency behavior in the
region around the last Coulomb peaks (Fig. 4) with a 50 GHz
microwave-signal applied to gate ${P_{L}}$. At the dotted line the
01 and 10 charge states are degenerate in energy, so one electron
can tunnel back and forth between the two dots. Away from this
line there is an energy difference and only one charge state is
stable. However, if the energy difference matches the photon
energy, the transition to the other dot is possible by absorption
of a single photon. Such photon-assisted tunneling events give
rise to the two lines indicated by the arrows. At the lower
(higher) line electrons are pumped from the the left (right) dot
to the other side, giving rise to a negative (positive)
photon-assisted current. We find that the distance between the
dotted line and the photon-assisted tunneling lines scales, as
expected, linearly with frequency \cite{wilfred}.

The realization of a controllable few-electron quantum dot circuit
represents a significant step towards controlling the coherent
properties of single electron spins in quantum dots \cite{lieven}.
Integration with the QPCs permits charge read-out of closed
quantum dots. This read-out can become single-shot, with the
charge detection started after the coherent manipulation of the
double dot quantum states. This procedure maximally reduces
backaction effects from the detector. Present experiments focus on
increasing the speed of the read-out such that the determination
of the charge state is faster than the mixing time, i.e. the time
in which the measurement introduces transitions between the charge
states \cite{delsing}.

We thank T. Fujisawa, T. Hayashi, Y. Hirayama, C. J. P. M.
Harmans, B. van der Enden, and R. Schouten for discussions and
help. This work was supported by the Specially Promoted Research,
Grant-in-Aid for Scientific Research, from the Ministry of
Education, Culture, Sports, Science and Technology in Japan, the
DARPA-QUIST program (DAAD19-01-1-0659), and the Dutch Organisation
for Fundamental Research on Matter (FOM).


\end{document}